\documentclass{article}
\usepackage[T1]{fontenc}
\usepackage[utf8]{inputenc}
\usepackage{ismir} 
\usepackage{amsmath,amssymb,cite,url}
\usepackage{graphicx}
\usepackage{color}
\usepackage{stfloats}
\usepackage{siunitx}
\usepackage{subcaption}
\usepackage{enumitem}

\usepackage{booktabs}

\usepackage{xcolor}

\definecolor{mylinkcolor}{RGB}{0,70,140}
\definecolor{mycitecolor}{RGB}{0,120,90}
\definecolor{myurlcolor}{RGB}{140,0,90}

\usepackage[
    bookmarks=false,
    colorlinks=true,
    linkcolor=mylinkcolor,
    citecolor=mycitecolor,
    urlcolor=myurlcolor
]{hyperref}

\newcommand{\DD}{\mathcal{D}}
\newcommand{\EE}{\mathcal{E}}
\newcommand{\RR}{\mathbb{R}}

\DeclareMathOperator*{\argmax}{arg\,max}
\DeclareMathOperator*{\argmin}{arg\,min}

\newif\ifarxiv

\arxivfalse  

\title{Steering dense music retrieval\\ with open-vocabulary concept discovery}

\multauthor
  {Julien Guinot$^{1,2}$ \hspace{1cm} Alain Riou$^1$ \hspace{1cm} Elio Quinton$^1$ \hspace{1cm} György Fazekas$^2$}
  {
  $^1$ Music \& Audio Machine Learning Lab, Universal Music Group, London, U.K.\\
  $^2$ Centre for Digital Music, Queen Mary University of London, U.K.\\
  {\tt\small j.guinot@qmul.ac.uk}
  }

\def\authorname{J. Guinot, A. Riou, E. Quinton and G. Fazekas}

\begin{document}

\maketitle

\begin{abstract}
Controllable music retrieval lets users find music that is, for example, \textit{more ambient}, \textit{less distorted}, or \textit{without guitar} while preserving the other semantic content of an original seed query. Sparse autoencoders (SAEs) are a promising interface for this kind of concept-level control, but a key problem remains: given a free-form text concept, which sparse features should be edited? In shared multimodal embedding spaces, standard attribution methods often select neurons that match the concept’s wording but not the audio examples that express it. This leads to weak or unstable edits: relevant features are missed when concepts are distributed across neurons, while others are selected due to text alignment rather than audio-side structure.

We address this with a lightweight, training-free method that recovers a sparse set of audio features whose decoded representation reconstructs the target concept while remaining consistent with audio-space geometry. This reframes concept attribution as a sparse inversion problem rather than a text-side neuron-ranking heuristic. The method requires neither paired audio--text supervision nor SAE retraining. We evaluate this approach in steerable music retrieval and show that the recovered supports align more closely with concept-bearing audio examples and achieve a stronger trade-off between edit strength and preservation than alignment baselines, enabling more precise concept amplification and suppression with reduced drift on preservation metrics.
\end{abstract}

\section{Introduction}\label{Section:Intro/Related}

Dense retrieval underpins modern music search and recommendation systems, where tracks are embedded into a shared space and retrieved by similarity to text or audio queries. For example, a user might search for “ambient electronic music” or retrieve tracks similar to a reference clip. While effective at capturing overall semantic similarity, these systems offer limited control over specific attributes—such as instrumentation, mood, or production characteristics—without altering the broader content of the query. This motivates \emph{steerable retrieval}: modifying the query embedding at retrieval time by amplifying or suppressing directions associated with a concept before nearest-neighbour search.

Sparse autoencoders (SAEs) provide a natural interface for this problem. By forcing each input to be reconstructed from only a small subset of features, sparsity encourages a more factorized representation in which active neurons tend to capture distinct directions of variation. This makes SAE features attractive as localized control directions for targeted retrieval.
Given a pretrained embedding space, an SAE learns a sparse code that reconstructs the original representation, decomposing dense embeddings into more interpretable units~\cite{jumprelu,bussmann2024batchtopk,bussmann2025learning,costa2025evaluating,gao2024scaling,lieberum2024gemma,pach2025sparse,zaigrajew2025interpreting}. This has made SAEs useful for mechanistic interpretability, but also as practical interfaces for analysing and manipulating learned representations. Recent work increasingly treats SAE features as controllable units. In language models, they have been used to steer generation \cite{makelov2024evaluating,chalnev2024improving,zhao2025denoising,wu2025interpreting}; in multimodal and generative systems, they support concept amplification, suppression, and style control \cite{pach2025sparse,kim2025concept}. SAEs have also been applied to dense retrieval embeddings, to interpret retrieval behaviour and to expose sparse latent spaces that can be manipulated post hoc through attributed concept neurons\cite{kang2025interpret,park2025decoding}. Together, these works suggest that SAEs can serve as actionable control interfaces for retrieval.

The central bottleneck is \emph{concept attribution}: given a free-form concept such as \emph{jazzy}, \emph{melancholic}, or \emph{female vocals}, which sparse features should be edited? Existing approaches rely on manual or example-based attribution: neuron inspection, exemplar mining, and post-hoc naming \cite{pach2025sparse,kang2025interpret,singh2025discovering,thasarathan2025universal,papadimitriou2025interpreting,makelov2024evaluating,park2025decoding,o2024disentangling} or on geometric probing, which matches concept text embeddings to the SAE decoder and selects the most aligned neurons \cite{kaushik2026decomposing,rao2024discover,dreyer2025attributing}. These methods are typically training-free but assume that a concept localizes to one individually aligned neuron. This assumption is increasingly questionable. Recent work shows that SAE features are not canonical units: concepts may be \emph{split} across multiple features, while features may \emph{absorb} several related factors \cite{leask2025sparse,chanin2024absorption,korznikov2025ortsae}. Good reconstruction and sparsity therefore do not guarantee faithful downstream control \cite{makelov2024evaluating}. The issue is sharper in multimodal spaces, where semantic structure is unevenly distributed across modalities and mixed with contextual or hub-like neurons \cite{zaigrajew2025interpreting,pach2025sparse}. Although paired text and audio are trained to be close, they do not occupy exactly the same regions of the shared space~\cite{MindTheGap}. A text embedding for \emph{electric guitar} may point toward the right semantic neighbourhood without coinciding with the audio embeddings of guitar examples. As a result, neurons that are most aligned with the concept on the text side may not be the neurons active for audio examples expressing that concept. For steerable retrieval, we therefore need sparse supports that reflect how the concept is represented in audio embeddings (\textit{audio-faithful}), not only how well each neuron matches the text query.


We study this problem in music retrieval and treat concept
attribution as sparse support recovery rather than neuron
ranking. Instead of selecting neurons independently by
text alignment, we solve a lightweight, training-free inverse problem: given a concept embedding, recover a sparse code whose decoded representation reconstructs the concept while remaining close to the empirical audio distribution. Figure~\ref{fig:main} summarizes the approach. We train SAEs on audio embeddings from joint music--text encoders, show that cosine-based probing recovers supports poorly aligned with concept-bearing audio examples, and replace this heuristic with sparse inversion under geometric and distributional constraints. We evaluate both attribution strategies for steerable retrieval, including concept amplification, suppression, and concept-specific comparison.

Our contributions are threefold:


\begin{figure*}[t!]
    \centering
    \begin{subfigure}[t]{0.32\textwidth}
        \centering
        \includegraphics[height=1.2in]{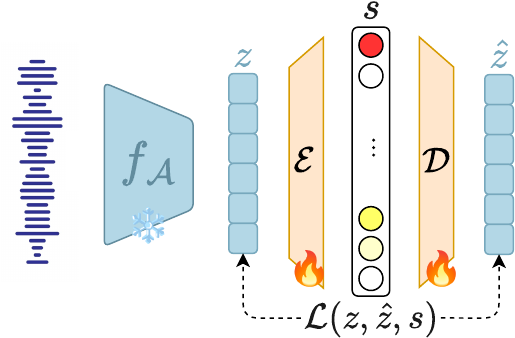}
        \caption{\textbf{Training}}
        \label{fig:main-training}
    \end{subfigure}%
    \hspace{0.01\textwidth}
    \begin{subfigure}[t]{0.32\textwidth}
        \centering
        \includegraphics[height=1.2in]{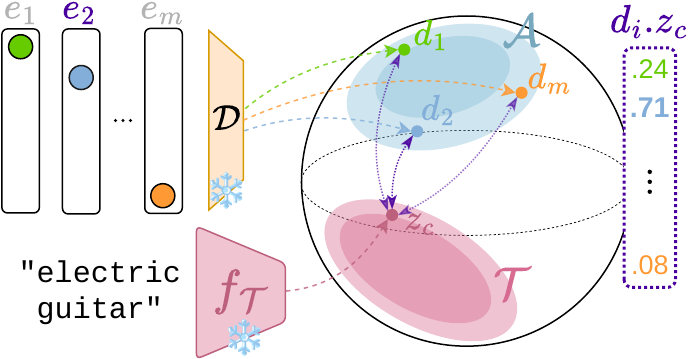}
        \caption{\textbf{Discover-Then-Name} cosine probing}
        \label{fig:main-dtn}
    \end{subfigure}
    \hfill
    \begin{subfigure}[t]{0.32\textwidth}
        \centering
        \includegraphics[height=1.2in]{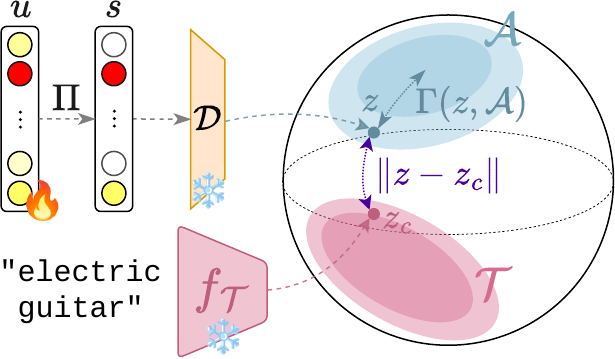}
        \caption{\textbf{Sparse Modality Inversion} (ours)}
        \label{fig:main-ours}
    \end{subfigure}
    \caption{
        Overview of the proposed approach.
        \textbf{(a)} Unimodal SAEs are trained on 10-second music snippet embeddings $z_a$ to decompose the dense embedding space into sparse codes $s$.
        \textbf{(b)} \textbf{Discover-Then-Name} consists in decoding one-hot vectors $e_1, \dots, e_m$ and picking the one, here {\color[HTML]{82AED9}$d_2$}, which aligns best with the concept embedding {\color[HTML]{BD607D}$z_c$}.
        \textbf{(c)} We instead derive concept support via \textbf{sparse inversion} by jointly optimizing the distance {\color[HTML]{4A009E}$\| z - z_c \|$} and the manifold regularization {\color[HTML]{628C9E}$\Gamma(z, \mathcal{A})$}.
    }
    \label{fig:main}
\end{figure*}

\begin{itemize}[leftmargin=*, itemsep=0.2em, topsep=0.2em, parsep=0pt, partopsep=0pt]
    \item We formulate controllable music retrieval as sparse feature intervention in SAE spaces and identify concept attribution as the key obstacle to open-vocabulary control.
    \item We show that standard attribution strategies are unreliable in multimodal settings: concept support is often distributed across multiple features, and text-side alignment does not recover audio-faithful sparse support.
    \item We introduce a lightweight, training-free inversion method for mapping free-form concepts to audio-side sparse supports, and show that it yields more precise, more manifold-faithful retrieval edits.
\end{itemize}

Code\footnote{\href{https://github.com/Pliploop/sparse-steer-retrieval/}{https://github.com/Pliploop/sparse-steer-retrieval/}} and examples are released. Additional examples are available on the companion website\footnote{\href{https://pliploop.github.io/sparse-steer-retrieval/}{https://pliploop.github.io/sparse-steer-retrieval/}}.

\section{Background}

\subsection{Sparse Autoencoders}

Let $f : \mathcal{X} \to \mathcal{A} \subset \RR^d$ be a pretrained representation learning model.
A Sparse Autoencoder (SAE) consists of a trainable encoder-decoder pair $(\EE, \DD)$, and a sparsity constraint $\Pi$ such that, for all embeddings $z = f(x) \in \mathcal{A}$:
\begin{equation*}
    u = \EE(z) \in \RR^m, \quad
    s = \Pi(u) \in \RR^m, \quad
    \hat{z} = \DD(s) \in \RR^d.
\end{equation*}
Typically $m\gg d$.
The training objective is:
\begin{equation}
    \mathcal{L}(z,\hat z,s) = \|z-\hat z\|_2^2 + \lambda\,\Omega(s),
\end{equation}
where $\Omega(s)$ is an optional sparsity regularizer, and $\lambda > 0$.


In practice, $\EE$ and $\DD$ are usually linear layers. SAE variants differ mainly in the sparsifying map $\Pi$, and sometimes in the optional regularizer $\Omega$. In vanilla/ReLU SAEs \cite{cunningham_sparse_2023}, $\Pi$ is a rectifying nonlinearity and $\Omega(s)=\|s\|_1$ encourages sparsity. In TopK SAEs \cite{gao2024scaling}, $\Pi$ keeps the $K$ largest activations per sample; in BatchTopK SAEs \cite{bussmann2024batchtopk}, it applies a $K$-related threshold at the batch level; and in Matching Pursuit SAEs \cite{costa2025evaluating}, it is a greedy pursuit procedure over the decoder dictionary $\DD$. Thus, for the TopK, BatchTopK, and Matching Pursuit variants considered here, sparsity is enforced by $\Pi$ rather than by an explicit $\Omega$ penalty. We write $L_0=\|s\|_0$ for the number of active coordinates.

\subsection{Discover-Then-Name Cosine Probing}\label{Subsection: Discover-Then-Name}

Many SAE concept-labeling approaches rely on post-hoc neuron naming. In joint multimodal models, a common method is to use text alignment directly.
Let $f_\mathcal{T}$ be the text encoder and $f_\mathcal{A}$ be the audio encoder of a multimodal embedding model such as CLAP \cite{clap}, and let $e_1, \dots, e_m$ be the canonical basis of $\RR^m$.

With \textit{Discover-Then-Name} (DTN)~\cite{rao2024discover}, each neuron $i \in \{1, \dots, m\}$ is projected to the audio manifold by computing $d_i = \DD(e_i)$. 
Then, text concepts $c$ are attributed to the neuron $i_c$ whose projection $d_{i_c}$ maximizes the cosine similarity with $z_c = f_{\mathcal{T}}(c)$, as shown in Figure~\ref{fig:main-dtn}:
\begin{equation}
  i_c = \argmax_i \cos(d_i, z_c).
\end{equation}



This rests on the assumption that text prompts can query an audio-trained dictionary because paired text and audio occupy aligned directions. DTN therefore selects the decoder atom closest to the concept text embedding. This is simple and training-free, but assumes that concepts localize to single sparse features and that text-side alignment identifies the same features used by corresponding audio examples. Both assumptions can fail: concepts may be distributed across neurons, and text and audio embeddings remain separated by a modality gap~\cite{MindTheGap,Shi2023,slap}. Multimodal SAEs can also exhibit split dictionaries with modality-specific features~\cite{papadimitriou2025interpreting,kaushik2026decomposing}, so text-aligned atoms may not be activated by concept-bearing audio. Thus, text alignment alone does not imply faithful audio-side support.



\begin{figure*}[t]
    \centering
    \def\conceptleftwidth{0.66}
    \def\conceptrightwidth{0.33}

    \begin{subfigure}[t]{\conceptleftwidth\linewidth}
        \vspace{0pt}
        \centering
        \includegraphics[width=\linewidth]{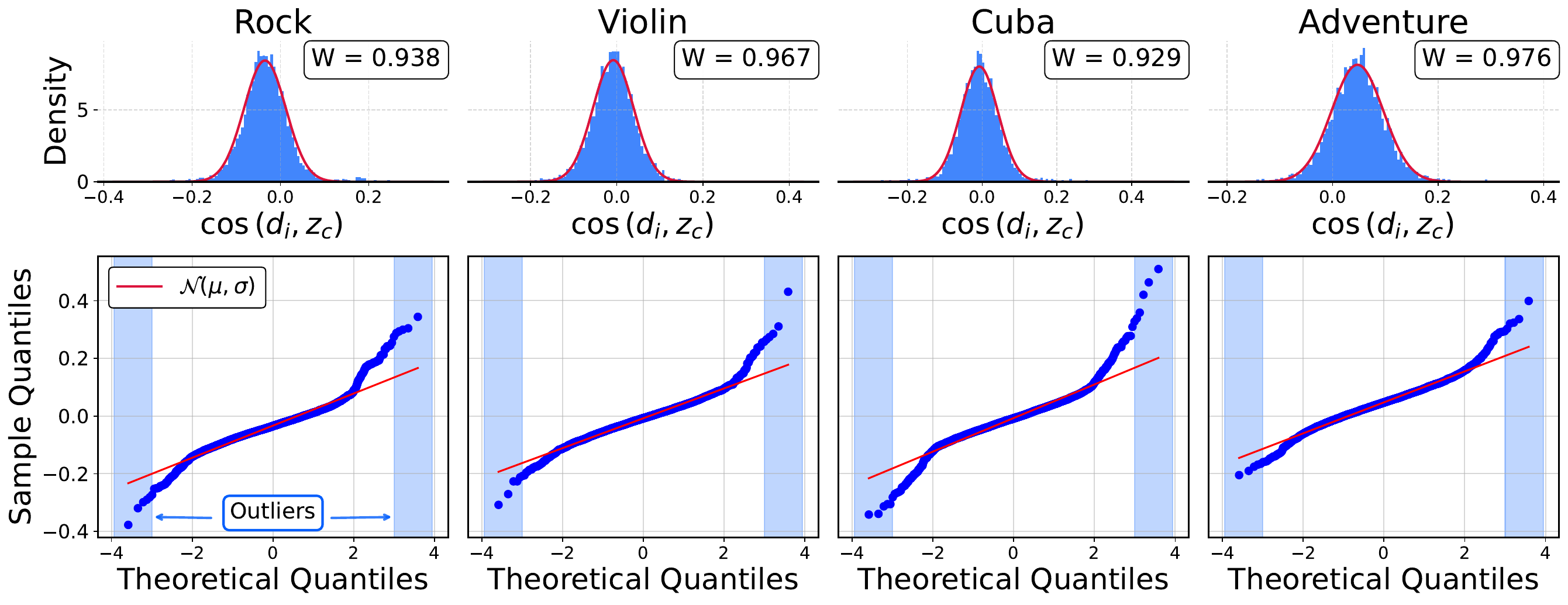}
        \caption{Cosine similarity distributions for \textbf{Rock}, \textbf{Violin}, \textbf{Cuba} and \textbf{Adventure}.}
        \label{fig:concepts4}
    \end{subfigure}
    \hfill
    \begin{subfigure}[t]{\conceptrightwidth\linewidth}
        \vspace{0pt}
        \centering
        \includegraphics[width=.78\linewidth]{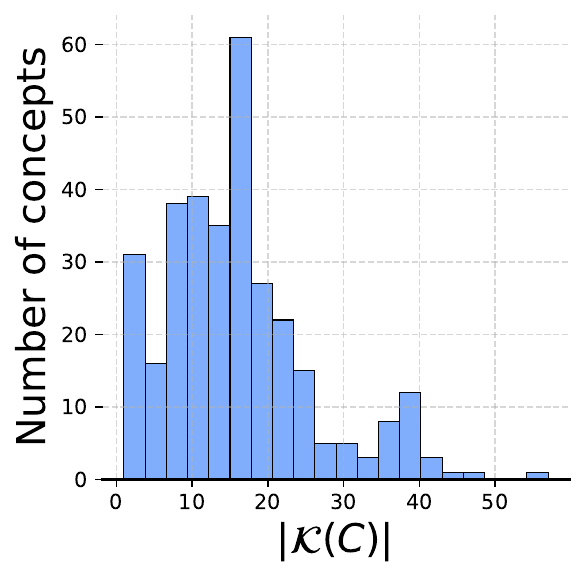}
        \caption{Concept cardinalities under $4\sigma$ filtering.}
        \label{fig:wrap}
    \end{subfigure}

    \caption{
        \textbf{(a)} Distribution of cosine similarity between $z_c$ and decoder atoms $d_i$ for selected concepts using a BatchTopK SAE trained on MuQ embeddings with $L_0 = 10$. Red line shows Fitted Gaussian and $W \in [0,1]$ is the Shapiro-Wilks test statistic for normality \cite{shapiro1965analysis}.
        \textbf{(b)} Histogram of concept cardinalities $|\mathcal{K}(c)|$ obtained by filtering cosine similarities outside the $4\sigma$ confidence interval, showing variable concept support sizes even under naive statistical filtering.
    }
    \label{fig:combined}
\end{figure*}

\section{Method and Observations}

\subsection{SAE Training}


We train SAEs on audio embeddings from JamendoMaxCaps \cite{roy2025jamendomaxcaps}. Audio clips are encoded with the audio towers $f_\mathcal{A}$ of CLAP \cite{clap} and MuQ \cite{zhu2025muq}; from each 10-second frame we extract an embedding $z_a \in \mathbb{R}^{512}$.
We train BatchTopK SAEs\footnote{We also experimented with TopK~\cite{gao2024scaling} and Matching Pursuit SAE~\cite{costa2025evaluating}, with similar results overall. Here, we focus on BatchTopK SAE, the most widely adopted SAE technique \cite{bussmann2024batchtopk}.} \cite{bussmann2024batchtopk} with $m = 4096$ latent features (expansion factor 8), sparsity levels $L_0 \in \{5,10,20,50,100\}$, and decoder $\DD$ instantiated as a linear layer with weights tied to the encoder. 
All models are trained for 50k steps using AdamW (learning rate $= 10^{-4}$) on a single L40 GPU.

\subsection{Concepts Appear as Neuron Bundles}

We first examine the decoder geometry induced by DTN
(Section~\ref{Subsection: Discover-Then-Name}). Using an internal music-tag taxonomy, we collect the top 50 tags in each of the categories
\emph{Genres}, \emph{Moods}, \emph{Instruments}, \emph{Usage}, and
\emph{Era}, yielding roughly 300 short-form concepts. For each concept
$c$ with text embedding $z_c$, we compute cosine similarities between
$z_c$ and every decoder atom $d_i$. We then fit a Gaussian to each
concept-specific similarity distribution and define a DTN support
$\mathcal{K}(c)$ by retaining atoms whose similarity exceeds the fitted
mean by more than four standard deviations. This conservative threshold
selects only strong positive outliers under the Gaussian hypothesis.

Figure~\ref{fig:combined} shows example similarity distributions and
Q-Q plots. The bulk of decoder similarities is close to Gaussian,
consistent with many weakly related atoms behaving like near-uniform
directions on the hypersphere. At the same time, the Q-Q plots exhibit
a shoulder of positive outliers: a small set of atoms that is more
aligned with the concept than the null model would predict. Crucially,
these outliers rarely reduce to a single standout neuron. DTN therefore already behaves like a bundle-selection problem rather than a one-neuron naming problem. Figure~\ref{fig:wrap} summarizes the resulting support cardinalities
$|\mathcal{K}(c)|$ across concepts. Even for low values of $L_0$,
multiple neurons often contribute to the same text-side concept, and
support size varies substantially across concepts. This challenges the
usual DTN interpretation in which a single neuron is assigned to a
concept by geometric alignment. Throughout the paper, $\mathcal{K}(c)$
denotes the weighted set of neurons attributed to concept $c$: under
DTN, it comes from cosine scores, while under our approach it is
recovered from the optimized sparse code
(Section~\ref{sec:sparse_inverse_dictionary_geometry}).




\subsection{Inversion for Concept Attribution}
\label{sec:sparse_inverse_dictionary_geometry}

We seek a concept-to-neuron mapping determined by geometry under the learned SAE dictionary rather than by marginal alignment statistics. A neuron can correlate with a concept without being required to reconstruct it. We therefore cast text-to-audio bridging as a sparse inverse problem in the audio SAE reconstruction space.

Let $\DD$ denote the linear decoder of the SAE trained on audio embeddings, 
and let $z_c \in \mathbb{R}^d$ be the concept text embedding.
Since audio and text embeddings lie on the hypersphere,
we optimize a latent pre-activation $u$ so that the decoded sparse code matches the direction of $z_c$ under cosine distance $ d_{\cos}(a,b) = 1 - \frac{a \cdot b}{\|a\|\,\|b\|}$:
\begin{equation}
u_c^\star
=
\argmin_u
 \quad d_{\cos}\Big(z_c, \DD(\Pi(u))\Big).
\label{eq:cosine_inverse_section}
\end{equation}

We optimize over $u$ to maintain native training-time sparsity as $\Pi$ is the same sparsifying operator used during SAE training. Unlike cosine probing, which ranks neurons independently, Eq.~\eqref{eq:cosine_inverse_section} selects neurons because they \emph{jointly} reconstruct the target concept direction under the audio dictionary. The resulting code is therefore a sparse explanation of the concept in the basis actually used for steering retrieval.

In practice, unconstrained inversion can drift towards text-side directions that do not lie on the audio manifold \cite{mistretta2025cross,dell2025training}. To bias the solution toward plausible audio embeddings, we add a distributional regularizer $\Gamma(\mathcal{A}, \DD(\Pi(u)))$, implemented as a Mahalanobis distance with respect to an empirical audio embedding distribution $\mathcal{A}$ estimated from 10k test-set MaxCaps audio embeddings.
Our full objective is therefore:
\begin{equation}
\begin{aligned}
u_c^\star
&=
\argmin_u \Big(
d_{\cos}\Big(z_c, \DD(\Pi(u))\Big)
+ \gamma \, \Gamma(\mathcal{A}, \mathcal{D}(\Pi(u)))
\Big),
\end{aligned}
\label{eq:full_inverse_section}
\end{equation}
where $\gamma > 0$ controls manifold adherence to prevent unconstrained drift.
An alternative to the $\Gamma$ term would be to optimize through the full audio encoder without setting any distributional prior, as in \cite{mistretta2025cross}.
However, this would make the inversion process prohibitively slow for interactive steerable retrieval ($\sim 20$s\,/\,inversion on CPU). We hence adopt the more lightweight Mahalanobis regularization proposed in \cite{dell2025training}, yielding runtimes compatible with real-world applications ($\sim 20$ms\,/\,inversion on CPU).


We optimize Eq.~\ref{eq:full_inverse_section}
via gradient descent
using Adam~\cite{Adam}, but observed convergence issues, which we solve by initializing $u$ from the sparse code of the nearest audio neighbour to the text embedding $z_c$.
Alternatively, taking advantage of the linearity of the decoder $\DD$, we propose optimizing Eq.~\ref{eq:full_inverse_section} using FISTA, a fast algorithm designed for solving linear inverse problems~\cite{FISTA}.
We refer to these variants as \textbf{Adam inversion} and \textbf{FISTA inversion}.
A hyperparameter sweep showed that both methods remain stable across a range of configurations, and we fix $\gamma = 10^{-4}$ in all subsequent experiments.

Finally, to guard against the influence of overly generic frequent neurons, we apply an IDF-weighted penalty after optimization \cite{kaushik2026decomposing, dreyer2025attributing} (for fairness of comparison with DTN probing, to which we apply the same IDF-weighing):

\begin{equation}
s_k \leftarrow  w_k s_k,
\qquad
w_k = \log \frac{N}{\mathrm{df}(k)+\epsilon},
\label{eq:idf_weights_section}
\end{equation}
where $\mathrm{df}(k)$ counts the number of audio samples for which neuron $k$ is active. Frequent neurons therefore incur stronger sparsity pressure. The final support is obtained by optimizing Eq.~\ref{eq:full_inverse_section}, applying the IDF weighting in Eq.~\ref{eq:idf_weights_section}, and taking the top active coordinates. After optimization, we use the same weighting to rank the recovered coordinates and define the concept support $\mathcal{K}(c)$ as the active neurons in the optimized and weighted sparse code.


This yields a training-free concept-to-support mapping: instead of ranking atoms independently by text similarity, inversion selects a sparse set of atoms that jointly reconstructs the concept while staying near the audio distribution. In the following experiments, we compare these supports against cardinality-matched DTN supports.

\section{Concept Attribution Experiments}

Do the neurons selected for a concept actually correspond to the sparse support used by audio examples that instantiate it? A useful concept-to-support mapping should satisfy two properties: it should overlap with the active supports of positive audio examples, and those restricted activations should retain enough signal to support simple concept detection. We therefore evaluate cosine probing and inverse support recovery under a shared protocol.

We use the MaxCaps validation split~\cite{roy2025jamendomaxcaps}, retain the top-$20$ tags by frequency, and restrict evaluation to concepts with at least $10$ positive examples. For an audio example $x$ with sparse code $s$, let $\mathcal{S}(x)=\{j : s_j \neq 0\}$ denote its active support. For each concept $c$ with text embedding $z_c$, we recover a concept support $\mathcal{K}(c)$ using either DTN cosine similarities or our inverse procedure, matching support cardinality to training-time sparsity.
\begin{figure}
    \includegraphics[width=0.97\linewidth]{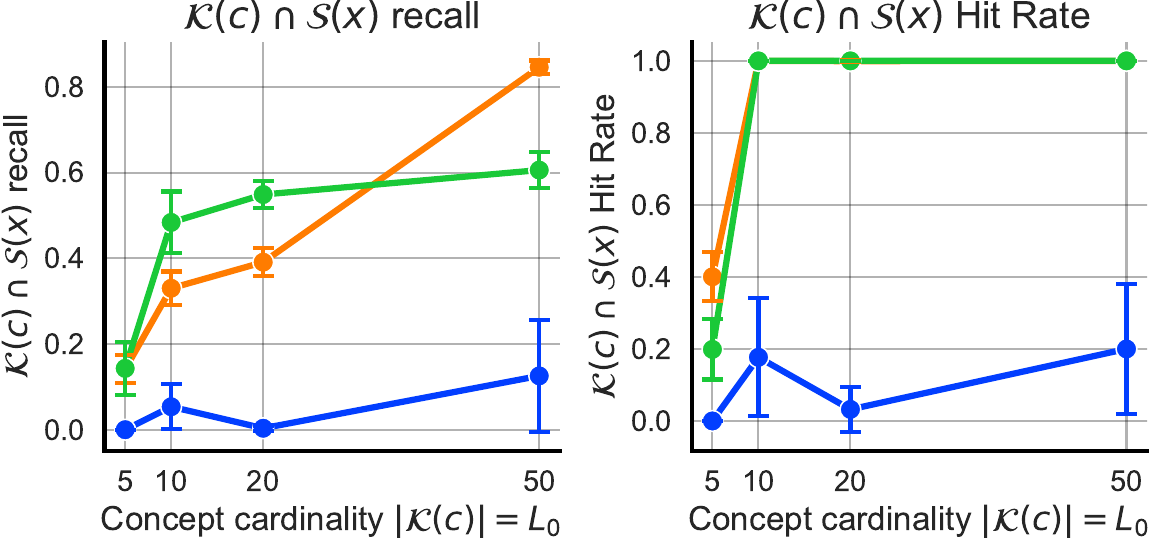}

    \vspace{8pt}
    
    \includegraphics[width=1\linewidth]{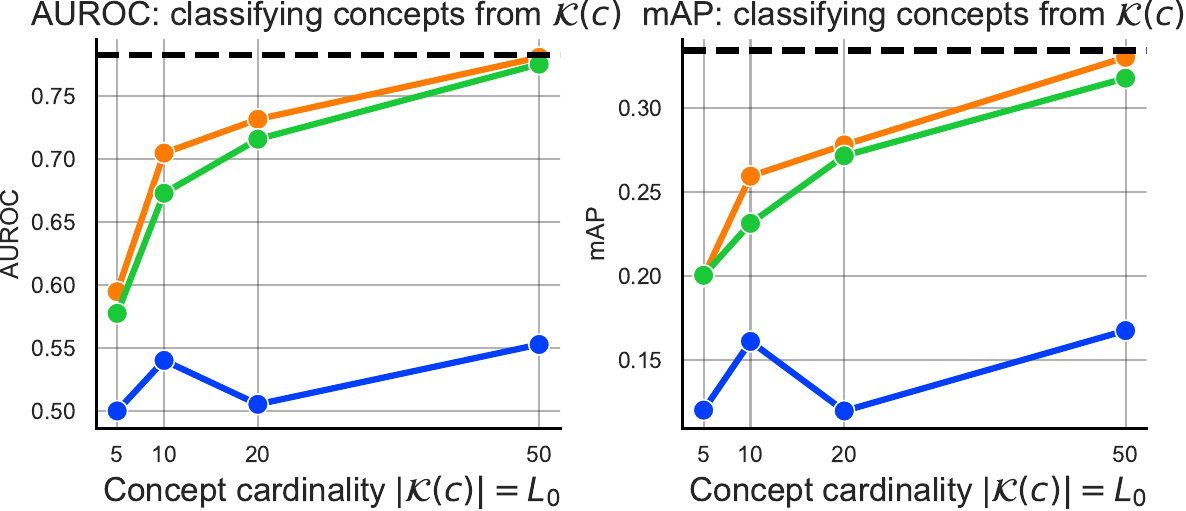}

    \vspace{6pt}

    \centering
    \includegraphics[width=0.95\linewidth]{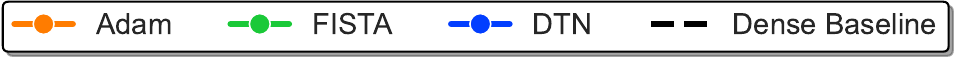}
    \caption{
        \textbf{Top:} Inverse $\mathcal{K}(c)$ align better with $\mathcal{S}(x)$ of $c$-tagged audio examples $x$ than cosine-matched baselines: inversion recovers faithful audio-side concept structure.
        \textbf{Bottom:} Inverse supports are more informative about concepts in audio embeddings than cosine-based baselines.
    }
    \label{fig:inverse_support_overlap}
\end{figure}


\subsection{Informativeness and overlap of concept supports}

\begin{figure*}[t]
    \hfill
    \includegraphics[width=0.97\linewidth]{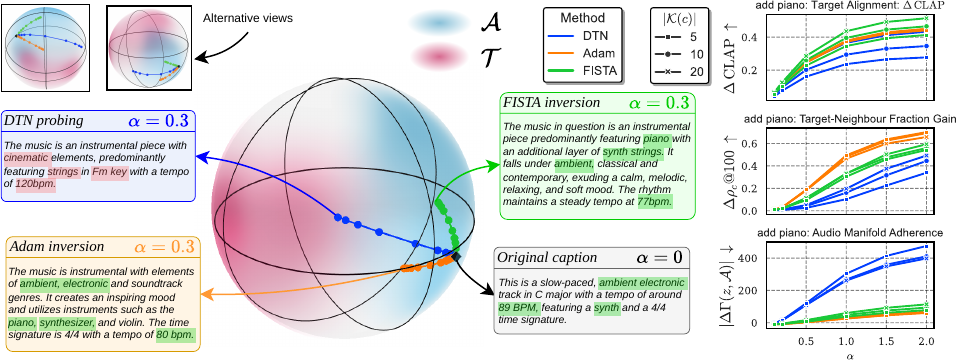}
    \caption{
    \textbf{Retrieving ambient electronic music with amplified ``piano''.}
    Starting from an audio embedding $z$, we navigate on the hypersphere by amplifying the concept ``piano'' with strength $\alpha$. 
    \textbf{Left:} PCA projections of the trajectories obtained with different methods as $\alpha$ increases.
    The original caption ($\alpha = 0$) and captions of nearest neighbour to the edited embeddings ($\alpha = 0.3$) are shown, with \colorbox[HTML]{BCF1C2}{correct} and \colorbox[HTML]{FDD4D4}{drifting} attributes highlighted.
    Sphere densities represent the {\color[HTML]{69E0F5}audio} ($\mathcal{A}$) and {\color[HTML]{F06EA2}text}  ($\mathcal{T}$) manifolds.
    \textbf{Right:} Evolution of the preservation and editing metrics as $\alpha$ increases, for $K \in \{5, 10, 20\}$.}
    \label{fig:piano}
    \vspace{-10pt}
\end{figure*}

We evaluate both families of supports in two complementary ways. First, we train a 1-hidden layer 512-wide binary classifier MLP on MaxCaps tags using only the activations indexed by $\mathcal{K}(c)$ and ask whether those restricted features predict the presence of concept $c$ in an audio example. This probe is intentionally weak: it is not meant to define a strong classifier, but to test whether the recovered neurons carry usable concept evidence.

Second, we compare $\mathcal{K}(c)$ directly against the empirical supports of positive examples $\mathcal{S}(x)$. We summarize overlap with mean bundle recall and bundle-hit rate. Concretely, bundle recall measures what fraction of the recovered bundle is active in positive examples, and $\mathrm{BundleHitRate}(c)$, the fraction of positive examples with nonzero overlap (see Figure \ref{fig:inverse_support_overlap}).
\vspace{-7pt}


\subsection{Results}
\label{subsec:structural_validation_overlap}


Figure~\ref{fig:inverse_support_overlap} shows that cosine-probed supports perform
poorly both structurally and functionally. Although the selected atoms
may look semantically plausible under decoder--text similarity, they
overlap weakly with positive audio supports, produce many zero-overlap
cases, and give weak concept discrimination. This suggests that cosine
probing often finds atoms that are near the concept in decoder space,
but not the sparse support used by concept-bearing audio examples.

Inverse supports close this gap. Under cardinality-matched comparisons,
they yield higher bundle hit rate, 10--70\% better bundle recall across
sparsities, and substantially better probe performance
($\sim$50\% AUROC $\rightarrow$ $\sim$65--80\%). Thus, inversion does not
only produce a more plausible concept bundle; it recovers supports that
are more often active in positive audio examples and more informative
for detecting the concept.

Taken together, these results provide a first structural validation of the inverse formulation. DTN appears to recover semantic association in decoder space, but not faithful support localization in the sparse audio representation. Inversion, by contrast, recovers supports that are more aligned with the neurons activated by positive examples and more informative for concept detection. This supports our central claim that concept attribution in multimodal sparse spaces should be treated as a geometric inverse problem rather than a marginal neuron-ranking heuristic.

\section{Sparse Steerable Retrieval}

We now test whether improved concept attribution also improves retrieval editing. The target use case is user-facing control such as ``\textit{I want similar songs without the guitar}'' or ``\textit{I want a track where the vocalist sounds more like this artist}.'' In this section, we ask whether sparse autoencoders support steerable music retrieval, and whether the better support alignment of inversion translates into more effective and more robust edits than DTN.

\subsection{Steering Retrieval}

We evaluate two steering operations: \textbf{concept suppression}, which reduces the presence of a concept in retrieved results, and \textbf{concept amplification}, which increases it. In this subsection, we treat tags as concepts. For each concept, we obtain a support either from DTN (cosine probing) or from one of the two inversion variants. In fixed-cardinality experiments, we use supports of size $K \in \{5,10,20\}$. For a concept support $\mathcal{K}(c)$ and edit strength $\alpha > 0$, amplification adds a weighted increment to every neuron in the bundle:\vspace{-5pt}
\begin{equation}
    s'(\alpha,\Pi,c) = s +\alpha\Pi(u_c^*)
\end{equation}
Concept suppression is defined analogously by subtracting from the bundle and clipping at zero:
\begin{equation}
    s'(\alpha,\Pi,c) = \max\bigl(0,\; s - \alpha\,\Pi(u_c^*)\bigr)
\end{equation}

We then decode the steered sparse activations back to the dense space through $\DD$ and $L_2$-normalize the edited embedding $z'(c,\Pi,\alpha)$ for retrieval. We $L_2$ normalize $\Pi(u_c^*)$ to act like directions rather than scale changes: support size controls \emph{which} coordinates are edited, while $\alpha$ remains the sole edit-strength parameter. Adding more neurons can therefore be interpreted as refining the edit direction. We expect the optimal support size to be concept-dependent (as hinted at by the different Q-Q outlier structures in Figure \ref{fig:combined}), and leave that trade-off to future work.

\subsection{Metrics}

We evaluate each edit along two axes: whether it changes the target
concept, and whether it preserves the rest of the query. Let $c$ be the
target concept, $z$ the original embedding, $z'$ the edited embedding,
and $\mathrm{NN}_{K}(z)$ the top-$K$ audio neighbours of $z$ in the
10k MaxCaps test set.

For target change, we use three metrics. First, we measure whether the
edited query retrieves more neighbours tagged with $c$:
\begin{equation}
    \Delta \rho_c@100 = \rho_c@100(z')-\rho_c@100(z),
\end{equation}
where $\rho_c@K(z)$ is the fraction of $\mathrm{NN}_{K}(z)$ tagged with
$c$. Second, we report the change in probability assigned to $c$ by an
audio tag classifier,
\vspace{-5pt}

\begin{equation}
    \Delta \hat{p}_{\mathrm{clf}}(c)
    =
    \hat{p}_{\mathrm{clf}}(c\mid z')-\hat{p}_{\mathrm{clf}}(c\mid z).
\end{equation}

Finally, as a text-alignment diagnostic, we report
$\Delta\mathrm{CLAP}=\cos(z',z_c)-\cos(z,z_c)$, where $z_c$ is the text
embedding of the concept.

For preservation, we measure whether the edit stays audio-like and
whether it changes unrelated tags. Audio-likeness is measured by $p_{\mathcal{A}}(z')$, from a held-out audio-vs-text classifier. Off-target drift is measured with a
co-occurrence-weighted Jaccard score over non-target neighbour tags,
reported as $\Delta J^{\mathrm{co}}_{\mathrm{other}}@100$. Tags that
naturally co-occur with $c$ are downweighted using
 dataset-level tag co-occurrence. This
avoids penalizing plausible changes such as adding \textit{electric
guitar} to a query and retrieving more \textit{rock}. We also
report $|\Delta\Gamma(z,\mathcal{A})|$, the change in Mahalanobis
distance to the empirical audio distribution, but treat it as a
diagnostic as the same distance is used in the inversion objective.
\vspace{-4pt}

\subsection{Results}

Figure~\ref{fig:piano} shows an example of amplifying the ``piano'' concept with $\alpha \in \{0.1,0.2,0.5,1.0,2.0\}$. All methods improve the target concept as $\alpha$ increases, but inversion does so more efficiently: at matched $\alpha$ and support size $K$, it moves the edited embedding closer to the concept while causing less preservation metric degradation. Both inverse variants also remain markedly closer to the audio manifold in Mahalanobis distance, suggesting that the recovered edit directions are better aligned with audio embeddings. In other words, DTN can produce sensible edits, but it often does so by moving toward the text manifold.

\begin{figure}[h]
    \centering
    \hspace{-5pt}
    \includegraphics[width=1\linewidth]{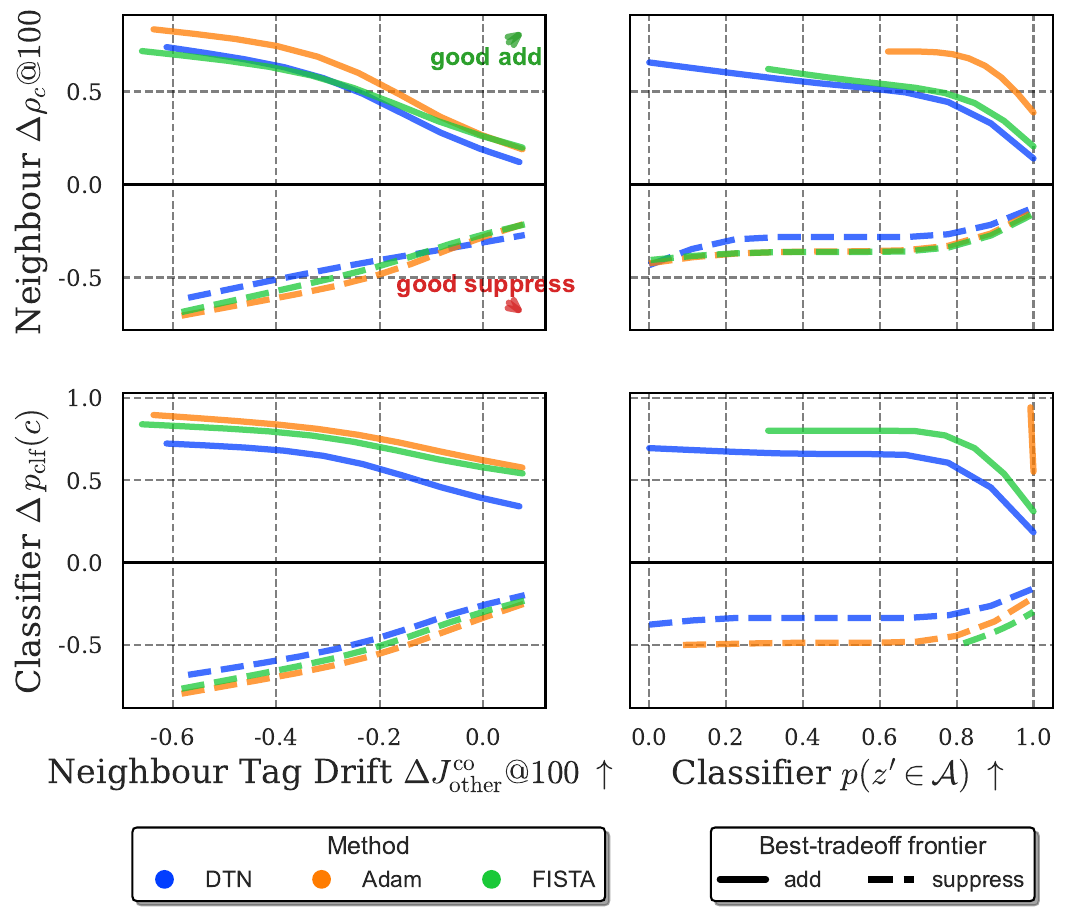}
    \caption{Edit--preservation Pareto frontiers for concept suppression and amplification show stronger trade-offs for inversion approaches than DTN cosine probing.}
    \label{fig:paretos}
    \vspace{-2pt}
\end{figure}
Figure~\ref{fig:paretos} shows edit--preservation trade-offs over the top 20 MaxCaps concepts. For each method, we collect all edit settings $(c,K,\alpha)$ and estimate an upper frontier using the $90^{\mathrm{th}}$ percentile edit score within bins of the preservation metric, retaining the top 10\% of outcome. We then apply a light smoothing spline only for visualization.

Across metrics, all methods exhibit the expected trade-off between stronger edits and greater preservation cost. The inverse methods occupy better high-score frontiers than cosine probing: at matched off-target drift or modality adherence, they achieve larger target-neighbour and classifier gains. This indicates that audio-side inverse supports are not only better aligned structurally, but translate into stronger retrieval edits at comparable preservation levels.
Poor regimes still exist for some $(c,K,\alpha)$ settings, suggesting limits from concept ambiguity, embedding-space quality, or SAE reconstruction fidelity.

\section{Limitations, Discussion, Conclusion}

Our method is constrained by the underlying text--music embedding space and SAE fidelity. If a concept is weakly grounded, rare, abstract, or diffusely represented across modalities, inversion may fail to recover reliable support, inheriting reconstruction error, feature splitting, and feature absorption. Performance also depends on sparsity and manifold-regularization hyperparameters, although the method is reasonably stable in practice. More fundamentally, SAEs applied directly to dense embeddings remain quasi-linear control interfaces. Future work could combine sparse factorization with more expressive or instance-adaptive retrieval mechanisms to enable smoother, more fine-grained slider-like edits.

We studied open-vocabulary concept attribution in sparse autoencoders for controllable music retrieval and showed that standard text-side heuristics fail due to cross-modal mismatch and feature splitting. Neurons aligned with a text concept often do not support it on the audio side, leading to weak and unstable edits. We instead reframe attribution as sparse support recovery, identifying neuron sets that jointly reconstruct the concept while remaining consistent with audio-space geometry. Across retrieval-editing experiments, inversion yields a stronger controllability--preservation trade-off than cosine probing while remaining lightweight, training-free, and applicable without paired supervision.

Overall, faithful concept control in multimodal SAEs depends less on identifying individually aligned neurons than on recovering the sparse support instantiating a concept in the target modality. This offers a practical route to interactive music retrieval, allowing users to refine results in real time through concept-level edits without retraining the embedding model or providing new examples.

\clearpage

\section{Acknowledgement}

This work is supported by the EPSRC UKRI Centre for
Doctoral Training in Artificial Intelligence and Music
(EP/S022694/1) and Universal Music Group.

\section{AI Usage Statement}

AI LLM assistants were used sparingly in the development of this project to accelerate the implementation of evaluation pipelines. AI assistance was also used sparingly in writing this paper to tighten phrasing, reword for concision, and smooth out mathematical notation consistency.

\bibliography{ISMIRtemplate}

\clearpage

\ifarxiv
\include{latex/Appendix}
\fi

\end{document}